\documentstyle[12pt,aps,prd,preprint]{revtex}
\begin{document}
\title{Bohr's Correspondence Principle and The Area Spectrum 
of Quantum Black Holes}
\author{Shahar Hod}
\address{The Racah Institute for Physics, The
Hebrew University, Jerusalem 91904, Israel}
\date{\today}
\maketitle

\begin{abstract}

During the last twenty-five years evidence has been 
mounting that a black-hole surface area has a {\it discrete} 
spectrum. Moreover, it is widely believed that 
area eigenvalues are {\it uniformally} spaced. There is,
however, no general agreement on the {\it spacing} of 
the levels. 
In this letter we use Bohr's correspondence principle to 
provide this missing link. 
We conclude that the area spacing of a black-hole is $4\hbar \ln 3$. 
This is the unique spacing 
consistent both with the area-entropy {\it thermodynamic} relation for 
black holes, with Boltzmann-Einstein formula in 
{\it statistical physics} and with {\it Bohr's correspondence principle}.

\end{abstract}
\bigskip

The necessity in a quantum theory of gravity was already 
recognized in
the 1930s. However, despite the flurry of activity on this
subject we still lack a complete theory of quantum gravity.
It is believed that black holes may play a major role in our 
attempts to shed some light
on the nature of a quantum theory of gravity (such as the role 
played by atoms in the
early development of quantum mechanics).

The quantization of black holes was proposed long ago in the
pioneering work of Bekenstein \cite{Beken1}. The idea was based on the
remarkable observation that the horizon area of nonextremal black
holes behaves as a classical {\it adiabatic invariant}. In the spirit
of Ehrenfest principle \cite{Ehren}, any classical adiabatic invariant
corresponds to a quantum entity with {\it discrete} spectrum,
Bekenstein conjectured that the horizon area of a non extremal quantum
black hole should have a discrete eigenvalue spectrum.

To elucidate the {\it spacing} of the area levels it is instructive to
use a semiclassical version of Christodoulou's reversible 
processes. 
Christodoulou \cite{Chris} 
showed that the assimilation of a neutral ({\it point}) particle by 
a (nonextremal) black hole
is reversible if it is injected at the {\it horizon} from 
a radial {\it turning point} of its motion. 
In this case the black-hole
surface area is left unchanged and the changes in the other black-hole parameters
(mass, charge and angular momentum) can be undone by another suitable
(reversible) process.
(This result was later generalized by Christodoulou and Ruffini 
for charged point particles \cite{ChrisRuff}).

However, as was pointed out by Bekenstein in his seminal
work \cite{Beken2} the limit of a {\it point} particle is not a legal
one in {\it quantum} theory. In other words, the particle cannot be
both at the horizon and at a turning point of its motion; this contradicts the {\it Heisenberg quantum uncertainty principle}.
As a concession to quantum theory
Bekenstein ascribes to the particle a {\it finite} proper radius $b$
while continuing to assume, in the spirit of Ehrenfest's theorem, that
the particle's center of mass follows a classical trajectory. 
Bekenstein \cite{Beken2} has shown that the assimilation of a 
finite size neutral particle inevitably causes an increase in the horizon area.
This increase is minimized if the particle is captured when its center
of mass is at a turning point a proper distance $b$ away
from the horizon \cite{Beken2}:

\begin{equation}\label{Eq1}
(\Delta A)_{\min}=8 \pi \mu b\  ,
\end{equation}
where $A$ is the black-hole surface area and $\mu$ is the 
rest mass of the particle.
For a point particle $b=0$ and one finds $\Delta A_{\min}=0$. 
This is Christodoulou's result for a reversible
process. However, a quantum particle is subjected to quantum
uncertainty. A relativistic quantum particle cannot be localized to
better than its Compton wavelength. Thus, $b$ can be no smaller than
$\hbar / \mu$. This yields a lower bound on the increase in
the black-hole surface area due to the assimilation of a (neutral)
test particle

\begin{equation}\label{Eq2}
(\Delta A)_{\min}=8 \pi {l_p}^2\  ,
\end{equation}
where $l_p=\left({G \over {c^3}}\right)^{1/2} {\hbar}^{1/2}$ is the
Planck length (we use gravitational units in which $G=c=1$).
It is easy to check that the reversible processes of Christodoulou and
Ruffini and the lower bound Eq. (\ref{Eq2}) of Bekenstein are valid
only for {\it non}-extremal black holes.  Thus, for nonextremal
black holes there is a {\it universal} (i.e., independent of the
black-hole parameters) minimum area increase as soon as one introduces
quantum nuances to the problem.

The universal lower bound Eq. (\ref{Eq2}) derived by Bekenstein is
valid only for {\it neutral} particles \cite{Beken2}.  Recently, Hod 
\cite{Hod} analyzed the capture of a quantum (finite
size) {\it charged} particle by a black hole and 
found a similar lower bound. 
The lower bound on the area increase 
caused by the assimilation of a charged particle is
given by \cite{Hod}

\begin{equation}\label{Eq3}
(\Delta A)_{\min}=4 {l_p}^2\  .
\end{equation}
As was noted by Bekenstein
\cite{Beken2} (for neutral particles) the underling physics 
which excludes a completely reversible process is the 
{\it Heisenberg quantum uncertainty principle}. 
However, for {\it charged}
particles it must be supplemented by another physical 
mechanism \cite{Hod} -- a Schwinger discharge of the black hole 
({\it vacuum polarization} effects). Without this physical mechanism one 
could have reached the reversible limit.

It is remarkable that the lower bound found for charged particles 
is 
of the same order of magnitude as the one given by
Bekenstein for neutral particles, even though they 
emerge from {\it different} physical mechanisms.
The {\it universality} of the fundamental lower bound (i.e., 
its independence on the black-hole
parameters) is clearly a strong evidence in favor of 
a {\it uniformly} spaced area spectrum for quantum
black holes (see Ref. \cite{Beken3}). Hence, one concludes that 
the quantization condition of the black-hole surface area 
should be of the form

\begin{equation}\label{Eq4}
A_n=\gamma {l_p}^2 \cdot n\ \ \ ;\ \ \ n=1,2,\ldots\ \  ,
\end{equation}
where $\gamma$ is a dimensionless constant.

It should be recognized that the precise values of the 
universal lower bounds Eqs. (\ref{Eq2}) and (\ref{Eq3}) can be
challenged. These lower bounds follow from the assumption that the
smallest possible radius of a particle is precisely equal to 
its Compton wavelength. Actually, the particle's size is not so
sharply defined. Nevertheless, it should be clear that 
the fundamental lower bound must be of the same order 
of magnitude as the one given by
Eq. (\ref{Eq3}); i.e., we must have $\gamma=O(4)$.
The small uncertainty in the value of $\gamma$ is the price we must
pay for not giving our problem a full quantum treatment.
In fact, the analyses presented in Refs. \cite{Beken2,Hod} 
are analogous to the well known semiclassical
determination of a lower bound on the ground state energy of 
the hydrogen atom \cite{Ehren}. Both analyses consider a 
{\it classical} object (an electron or a test particle) 
subjected to the Heisenberg 
uncertainty principle. The analogy with 
usual quantum physics suggests the next step -- 
a {\it wave} analysis of black-hole perturbations.

The evolution of small perturbations of a black hole 
are governed by a
one-dimensional wave equation. This equation was first derived by
Regge and Wheeler for perturbations of the Schwarzschild black hole
\cite{RegWheel}.  Furthermore, it was noted that, at late times, all
perturbations are radiated away in a manner reminiscent of the last
pure dying tones of a ringing 
bell \cite{Cruz,Vish,Davis}. To describe
these free oscillations of the black hole the notion of quasinormal
modes was introduced \cite{Press}. The quasinormal mode 
frequencies (ringing frequencies) are characteristic of the black hole itself.

The perturbation fields outside the black hole 
are governed by a one-dimensional 
Schr\"odinger-like wave equation (assuming a time dependence of the
form $e^{-iwt}$):

\begin{equation}\label{Eq5}
{{d^2 \Psi} \over {{d{r_*}}^2}} + \left [w^2 - V(r) 
\right]\Psi=0\  ,
\end{equation}
where the tortoise radial coordinate $r_*$ is related to the spatial
radius $r$ by
$dr_*=dr/ \left( 1-{{2M} \over r} \right)$ and the effective
potential is given by

\begin{equation}\label{Eq6}
V(r)=\left(1-{{2M} \over r} \right) \left[ {{l(l+1)} \over {r^2}}
  +{\sigma \over {r^3}} \right]\  ,
\end{equation}
where $M$ is the black-hole mass, $l$ is the 
multipole moment index, and
$\sigma=2, 0, -6$ for scalar, electromagnetic, and gravitational
perturbations, respectively.

The black hole's free oscillations (quasinormal modes) correspond to
solutions of the wave equation (\ref{Eq5}) with the physical boundary
conditions of purely outgoing waves at spatial infinity ($r_* \to
\infty$) and purely ingoing waves crossing the event horizon ($r_* \to
-\infty$) \cite{Detwe}.  The quasinormal modes are related to 
the pole singularities of the scattering amplitude in
the black-hole background. The ringing frequencies are located 
in the complex frequency plane characterized by Im($w$)$< 0$.  It turns
out that for a given $l$ there exist an 
infinite number of quasinormal
modes for $n=0, 1, 2,...$ characterizing modes with decreasing
relaxation times (increasing imaginary part) \cite{Leaver,Bach}. On 
the other hand, the real part of the frequency approaches a constant
value as $n$ is increased.

Our analysis is based on {\it Bohr's correspondence principle} (1923):
``transition frequencies at large quantum numbers should equal
classical oscillation frequencies''. Hence, we are interested in the
asymptotic behavior (i.e., the $n \to \infty$ limit) of the ringing
frequencies.  These are the highly damped black-hole oscillations 
frequencies, which are compatible with the statement (see, for
example, \cite{BekMuk}) ``quantum transitions do not take time'' 
(let $w=w_R -iw_I$, 
then $\tau \equiv {w_I}^{-1}$ is the effective 
relaxation time for the black hole to return to a quiescent
state. Hence, the relaxation time $\tau$ is 
arbitrarily small as $n \to \infty$).

The determination of the highly damped quasinormal 
mode frequencies of
a black hole is not a simple task. This is a direct 
consequence of an exponential
divergence of the quasinormal mode eigenfunctions 
at $r_* \to \infty$.
In fact, the asymptotic behavior of the ringing
frequencies is known only for the simplest case of a Schwarzschild
black hole.  Nollert \cite{Nollert} found that the asymptotic
behavior of the ringing frequencies of a Schwarzschild black hole is
given by

\begin{equation}\label{Eq7}
Mw_n=0.0437123-{i \over 4} \left(n+{1 \over 2} \right)
+O\left[(n+1)^{-{1/2}} \right]\  .
\end{equation}
It is important to note that the highly damped ringing frequencies
depends only upon the black-hole mass and is {\it independent} of $l$
and $\sigma$. This is a crucial feature, which is consistent with the
interpretation of the highly damped ringing frequencies (in the $n \gg
1$ limit) as characteristics of the {\it black hole} itself.  The asymptotic behavior Eq. (\ref{Eq7}) was later
verified by Andersson \cite{Andersson} using an 
independent analysis. 

We note that the numerical 
limit $Re(w_n) \to 0.0437123 M^{-1}$ (as $n \to \infty$)
agrees (to the available data given in \cite{Nollert}) with the 
expression $\ln 3/(8 \pi)$. This identification is supported by
thermodynamic and statistical physics arguments discussed below.
Using the relations $A=16 \pi M^2$ and $dM=E=\hbar w$ one 
finds $\Delta A=4{l_p}^2 \ln 3$. Thus, we conclude that the
dimensionless constant $\gamma$ appearing in Eq. (\ref{Eq4}) 
is $\gamma =4 \ln 3$ and the area
spectrum for the quantum Schwarzschild black hole is given by
 
\begin{equation}\label{Eq8}
A_n=4 {l_p}^2 \ln 3 \cdot n\ \ \ ;\ \ \ n=1,2,\ldots\ \  . 
\end{equation}

This result is remarkable from a {\it statistical physics} point of
view !  The semiclassical versions of
Christodoulou's reversible processes Refs. \cite{Beken2,Hod}, which
naturally lead to the conjectured area spectrum
Eq. (\ref{Eq4}), are at the level of mechanics, not statistical
physics. In other words, these arguments did not relay in any way on
the well known thermodynamic relation between black-hole surface area
and entropy.  In the spirit of Boltzmann-Einstein formula in
statistical physics, Mukhanov and Bekenstein \cite{Muk,BekMuk,Beken3}
relate $g_n \equiv exp[S_{BH}(n)]$ to the number of microstates of the
black hole that correspond to a particular external macrostate
($S_{BH}$ being the black-hole entropy).  In other words, $g_n$ is the
degeneracy of the $n$th area eigenvalue.  The accepted thermodynamic
relation between black-hole surface area and entropy \cite{Beken2} can
be met with the requirement that $g_n$ has to be an integer for every
$n$ only when
 
\begin{equation}\label{Eq9}
\gamma =4 \ln k\ \ \ ;\ \ \ k=1,2,\ldots\ \  .
\end{equation}
Thus, statistical physics arguments force 
the dimensionless constant $\gamma$ in Eq. (\ref{Eq4}) to 
be of the form Eq. (\ref{Eq9}). Still, a specific value of $k$
requires further input, which was not aveliable so far. This 
letter provides a first independent derivation of 
the value of $k$.  It should be
mentioned that following the pioneering work of Bekenstein
\cite{Beken1} a number of independent calculations (most of them in
the last few years) have recovered the uniformally spaced area
spectrum Eq.  (\ref{Eq4})
\cite{Kogan,Maggiore,Lousto,Peleg,LoukoMakela,BarKun,Makela,Kastrup}.
However, there is no general agreement on the spacing of the levels.
Moreover, {\it non} of these calculations is compatible with the
relation $\gamma =4 \ln k$, which is a direct consequence of the
accepted thermodynamic relation between black-hole surface area and
entropy. The relation $\gamma =4 \ln 3$ derived in this letter is 
the {\it only} one consistent both with the area-entropy
thermodynamic relation, with statistical physics 
arguments, and with Bohr's correspondence principle.

The universality of black-hole entropy (i.e., its direct
thermodynamic relation to black-hole 
surface area) and the universality of the lower bounds Eqs. 
(\ref{Eq2}) and (\ref{Eq3}) (i.e., their independence of the
black-hole parameters) suggest that the area 
spectrum Eq. (\ref{Eq8})
should be valid for a generic Kerr-Newman black hole.
Moreover, our analysis leads to a natural 
conjecture on the asymptotic behavior of the
highly damped quasinormal modes of a generic Kerr-Newman black hole.
Using the first law of black-hole thermodynamics

\begin{equation}\label{Eq10}
dM=\Theta d A + \Omega dJ\  ,
\end{equation}
where $\Theta={1 \over 4}(r_{+}-r_{-})/A$ and 
$\Omega=4 \pi a /A$ [$r_{\pm} =M \pm (M^2-a^2-Q^2)^{1/2}$ 
are the black hole's (event and inner) horizons and 
$a=J/M$ is the black-hole angular momentum per
unit mass] one finds

\begin{equation}\label{Eq11}
Re(w_n) \to 4 \Theta \ln 3+\Omega m\  ,
\end{equation}
as $n \to \infty$, 
where $m$ is the azimuthal eigenvalue of the field.
The asymptotic behavior of the Kerr-Newman ringing
frequencies was not determined directly so far. 
This is a direct consequence of
the numerical complexity of the problem. 
It is of great interest to compare 
the conjectured asymptotic behavior given in this letter 
with the results of direct numerical computations.
  
In summary, using a semiclassical version of 
Christodoulou's reversible 
processes (see Refs. \cite{Beken2,Hod}) 
one can derive a fundamental lower bound 
on the increase in black-hole surface area. 
The {\it universality} of the 
fundamental lower bound (i.e., 
its independence of the black-hole
parameters) is a strong evidence in favor of 
a {\it uniformly} spaced area spectrum for quantum
black holes. However, the spacing between 
area eigenvalues cannot be determined to better 
than an order of magnitude (the results 
presented in Ref. \cite{Hod} suggest that the area spacing is of
order $4 {l_p}^2$). 
This is a direct consequence of the semiclassical 
nature of these analyses. An analogy with 
usual quantum physics suggests the next step -- 
a {\it wave} analysis of black-hole perturbations. Applying 
Bohr's Correspondence Principle to 
the ringing frequencies which characterize a black hole, we derive
the missing link. We find the area spacing to be $4 {l_p}^2 \ln 3$,
which is in excellent agreement with the value predicted by the 
semiclassical analysis \cite{Hod}.
Moreover, this result is remarkable from a {\it
  statistical physics} point of view. 
The area spacing $4 {l_p}^2 \ln 3$ derived in this letter is 
the {\it unique} value consistent both with the area-entropy
{\it thermodynamic} relation, with {\it statistical physics} 
arguments (namely, with the Boltzmann-Einstein formula), 
and with {\it Bohr's correspondence principle}.

\bigskip
\noindent
{\bf ACKNOWLEDGMENTS}
\bigskip

I thank Avraham E. Mayo for helpful discussions.
This research was supported by a grant from the Israel Science Foundation.

\end{document}